\documentclass[namedreferences]{kluwer}
\usepackage{psfig}
\begin{document}
\begin{article}
\begin{opening}
\title{Structure of early-type galaxies: 2D fit of the light distribution
for a complete volume-limited sample} 

\author{E. \surname{Iodice}\email{iodice@sissa.it}}
\institute{International School for Advanced Studies, Trieste - Italy}
\author{M. \surname{D'Onofrio}\email{donofrio@pd.astro.it}}
\institute{Dipartimento di Astronomia - Universit\'a di Padova - Italy}
\author{M. \surname{Capaccioli}\email{capaccioli@na.astro.it}}
\institute{Dipartimento di Scienze Fisiche - Universit\`a di Napoli {\it
Federico II\/} - Italy\
Osservatorio Astronomico di Capodimonte, Napoli - Italy}

\runningauthor{Iodice et al.}
\begin{ao}
iodice@sissa.it
\end{ao}

\begin{abstract}
We outline the results of a two-dimensional (2D) fit of the light distribution
of early-type galaxies belonging to a complete volume-limited sample, and
briefly discuss the significant correlations among the structural parameters.
In particular we reconfirm that the lack of structural homology is likely a
characteristic of hot stellar systems.
\end{abstract}

\end{opening}

\section{Introduction}
The luminosity distribution of ellipticals and bulges of S0 galaxies has been
modelled for long using the $r^{1/4}$ formula (de Vaucouleurs 1948), which is
characterized by just two scale parameters, the effective radius $R_{e}$ and the
effective surface brightness $\mu_{e}$ of the isophote encircling half of 
the total light.
More recently \inlinecite{caon93} showed that a generalized de Vaucouleurs
formula, $r^{1/n}$, obtained by adding a free exponent $n$ \cite{sersic}, not
only gives a better fit to the light profiles (which is expected in view of 
the larger number of parameters), but it provides also an interesting
correlation between shape ($n$) and total luminosity ($L\sim I_e R^2_e$),
which in turn suggests that the family of elliptical galaxies is not
homologous.

The main criticism moved to this result is that 
the observed trend of $n$ with $R_e$ may be due to the presence of
hidden disks embedded in elliptical galaxies: the larger is this disk, 
the smaller comes the value of $n$.

Caon et al. (1993) had obtained their result by fitting the main-axes light
profiles using just the $r^{1/n}$ law. In view of the above criticism, we
decided to pinpoint the question by adopting a 2D fitting procedure allowing
the presence of the disk; 2D is needed in order to strongly couple and
exploit the different projection properties of components such as disk and
bulge which are expected to have quite different intrinsic thickness.

A non-linear least square fit minimizing the $\chi^2$ has been applied to the
same volume-limited sample of Caon et al. (1993): it contains $\sim 80$
early-type galaxies belonging to either the Virgo or the Fornax cluster. This
sample is $\sim 80\%$ complete down to $B_T\sim14$ mag, equivalent to $M_B
\sim -17.3$ if we assume a distance modulus $\Delta\mu = 31.3$ for both
clusters.

In the following we outline the adopted methodology and summarize
the results of this work (D'Onofrio et al., in preparation).

\section{The 2D fit}
Our 2D fitting models are based on the superposition of up to two components,
each one characterized by concentric and coaxial elliptical isophotes with
constant flattening. One such component may be thought as the projection of a
spheroid with finite intrinsic thichness, thus mimicking a bulge. For this
component only we allow a moderate degree of boxiness/diskiness. The other
component is thought to be a disk.

{\bf Model 1} is the 2D extension of the one used in their work by Caon et al.
(1993).
It consists of a spheroidal component only whose projected light distribution
follows the generalized de Vaucouleurs law:

\begin{equation}
\mu_{b}(x,y)=\mu_{e}+k \left[\left(\frac{r_{b}}{r_{e}}\right)^{1/n}-1 \right]
\end{equation}
with $k=2.5(0.068n-0.142)$ and
$r_{b}=\left[{x^{c}+\frac{\displaystyle y^{c}}{\displaystyle
(b/a)_b^{c}}}\right]^{1/c}$; $c$ is the parameter accounting for some
boxiness/diskiness ($c>2$ for boxy isophotes and $c<2$ for disky isophotes).

\medskip
{\bf Model 2} is made by the superposition of a bulge-like component with a
$r^{1/4}$ light distribution (same as eq. 1, with $n=4$) to an exponential
disk represented as:

\begin{equation}
\mu_{d}(x,y)=\mu_{0}+1.086\left(\frac{r_{d}}{h}\right)
\end{equation}
with $r_{d}=\left[{x^{2}+\frac{\displaystyle y^{2}}
{\displaystyle (b/a)_d^{2}}}\right]^{1/2}$.

\medskip
{\bf Model 3} is a generalization of model 2; the light distribution of
the bulge follows a $r^{1/n}$ law, but the disk remains exponential.
\bigskip

In summary, the structural parameters involved in these models are: the
effective surface brightness $\mu_{e}$ and the effective radius $R_{e}$ of the
spheroid, the free exponent $n$ when considered, the central surface
brightness $\mu_{0}$ and the scale length of the disk $h$ if included, the
apparent axial ratios $b/a$ for the bulge and the disk components, and 
the boxiness/diskiness parameter $c$.

\begin{figure}
\centerline{
\psfig{file=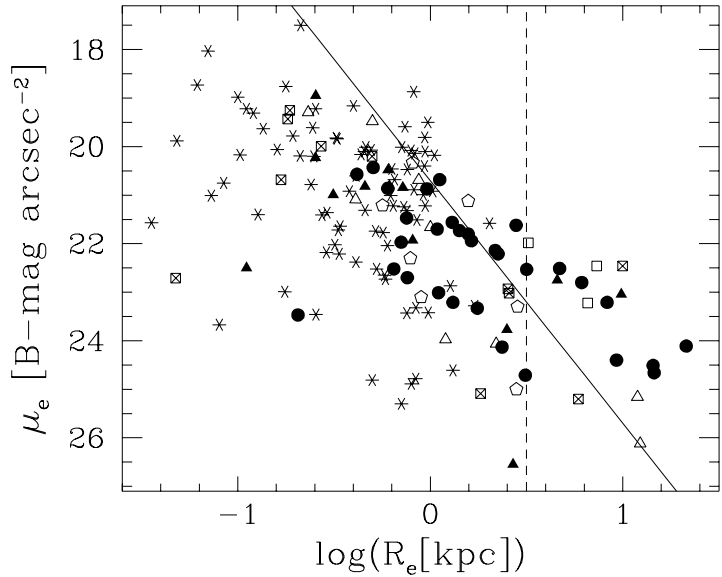,width=24pc,angle=-90}}
\caption{\small {
Plot of the effective surface brightness $\mu_{e}$ versus the
logarithmic effective radius $\log(R_{e})$ for the bulge component.
The bulges of spiral galaxies studied by de Jong (1996) 
are marked by asterisks. 
Filled circles and triangles mark the best fit for E and S0 galaxies, open
pentagons and triangles the fair fits, open squares and open squares with a
cross the poor fits.
The solid line represents a constant absolute magnitude 
$M_{B}=-19.3$.}}
\end{figure}

\begin{figure}
\centerline{
\psfig{file=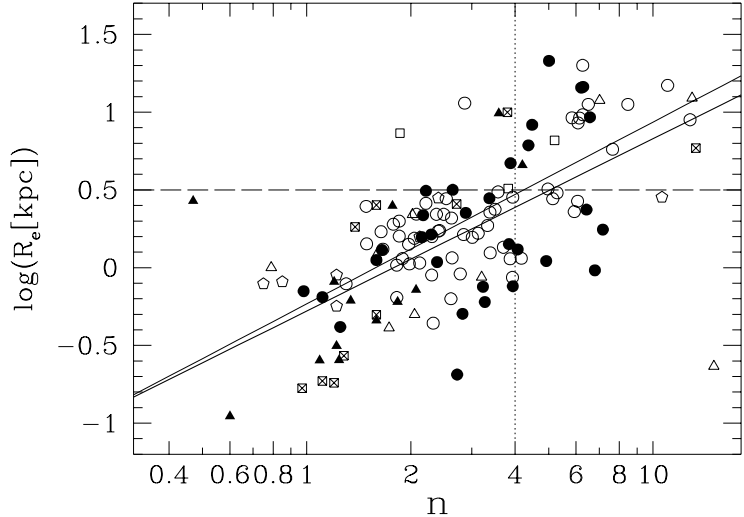,width=24pc,angle=-90}}
\caption{\small {Plot of the effective radius $\log(R_e)$  of the bulge
component versus the exponent $n$ of the generalized de Vaucouleurs law
(symbols as in Fig.1). Open circles represent the data
from Caon et al. 1993.}} 
\end{figure}

A number of simulations on artificial galaxies have been performed in order to
test the tool and to unveil the systematic effects that influence the model
parameters; in particular the seeing, which produces strongly biased results
even after proper convolution of the model, and can be taken care only by
masking out the central area of the galaxies. The simulations also provided a
more correct estimate of the errors of the model parameters.

\section{Result of the fit}
The three adopted models do not provide the same accuracy in reproducing the
light distribution of early-type galaxies. By examining the reduced $\chi^2$,
the 2D residuals, and by comparing the geometry of the model (ellipticity and
$a_{4}$ profiles) with the real galaxies, we concluded that the $r^{1/n} +
exp$ model has to be preferred to either the simple $r^{1/n}$ and the $r^{1/4}
+ exp$ models. Model 1 is comparable to model 3 in reproducing dwarf
elliptical galaxies ($M_B>-17$). Model 2 often produces extended luminous
halos not observed in the real objects.

It is evident that the reduced $\chi^2$ by itself does not guarantee the
physical significance of the best-fit solution for models with different
numbers of parameters. We performed also an F-test of the best-fit $\chi^2$
distributions, in order to verify whether the variances are different even
from a statistical point of view. Actually, however, we may try to attach a
physical meaning to the most complex and best-fitting model, that of $r^{1/n}
+ exp$, only after analyzing and falsifying the correlations among the
different parameters.

\begin{figure}
\centerline{
\psfig{file=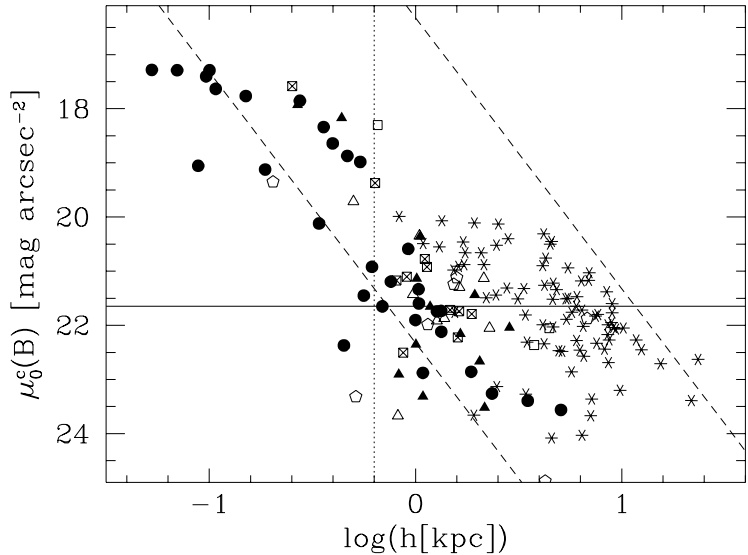,width=24pc,angle=-90}}
\caption{\small 
{Plot of the central corrected surface brightness $\mu_0^c$ versus the scale
length $h$ of the disks (symbols as in Fig.1). The solid line is the Freeman
value of constant $\mu_0^c(B)=20.75$. The data seem to follow the lines of
constant absolute magnitude (dashed lines).}}
\end{figure}

In general the best fitting model does not work in the inner seeing-dominated
region ($<1''$) and in the outer parts, where an extended luminous `halo' is
often found, but it is good in the intermediate regions for a large range of
$\mu$.
 
\section{Analysis of the bulge parameters}
We confirm the correlation between the effective radius $R_{e}$ and the
effective surface brightness $\mu_{e}$. The $\mu_{e} - \log(R_{e})$ relation
(Fig. 1) shows a family of large bulges with $R_{e}>3$ kpc and with low surface
brightness $\mu_{e}$, as found by \inlinecite{cap92}. The separation between
{\it ordinary} and {\it bright} galaxies is well visible, even if the gap
between the two families is less clear. The bulges of spirals all belong to
the ordinary family ($R_{e} <3$ kpc).

Fig. 2 shows the $n-\log(R_{e})$ relation.
The open circles represent the old distribution, found by \inlinecite{caon93}
for the same sample of galaxies by fitting the equivalent light profiles.
The dashed line marks the value $n=4$ of the de Vaucouleurs law.
By comparing the old relation with the new one, we observe 
that there is not a variation within the errors of the parameters.

\begin{figure}
\centerline{
\psfig{file=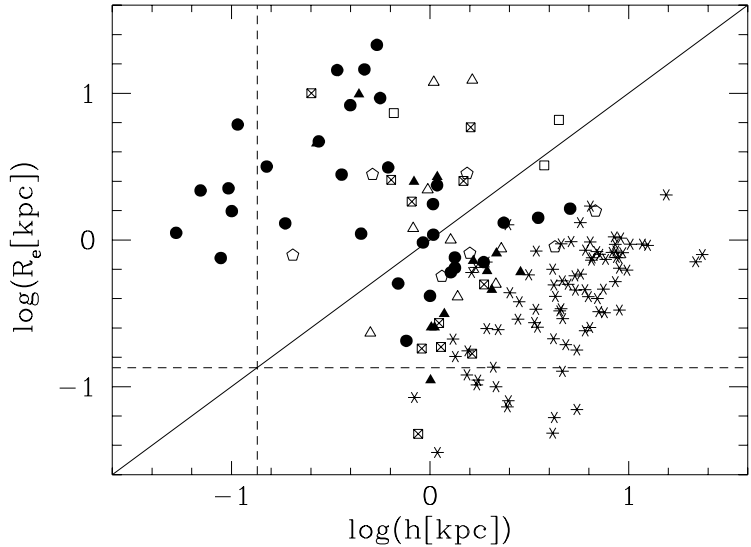,width=24pc,angle=-90}}
\caption{\small 
{Logarithmic plot of the scale length $R_e$ of the bulge
versus the the scale length $h$ of the disk components (symbols as
in Fig. 1). The dashed lines mark the seeing dominated regions.}}
\end{figure}

The $n$ parameter spans a wide range of values ($1<n<10$): low values
are typical of exponential bulges (already known to exist in spiral galaxies),
while high values are found in bright and large galaxies.

This result confirms that the different values of $n$ are not due to the
presence of a disk, but that there is a clear trend in the values of $n$ with
the galaxy size (and therefore with the total luminosity) for the
spheroidal components. It follows that the non-homology is likely a
characteristic of hot stellar systems.

\section{Analysis of the disk parameters}
Our best fit solutions tell us that in quite a few ($\sim 25\%$) elliptical
galaxies there is a central exponential component. This is not surprising for
the disky ellipticals, in which an `embedded disk', generally smaller and
brighter than normal disks in S0's and spirals, gives a characteristic disky
shape to the isophotes (\opencite{cap90}, \opencite{scorza95}).
However, we found an inner exponential component (with bright $\mu_0$ and
small $h$) also in the bright, boxy ellipticals.

We verified {\it a posteriori\/} that such small and bright `disks' are a first
order approximation of the inner cores of elliptical galaxies, well described
by the 'Nuker law' (\inlinecite{lauer}, \inlinecite{byun} and
\inlinecite{faber}).

Fig.3 shows the central surface brightness $\mu_{0}^{c}$ (corrected for 
inclination assuming a zero-thickness disk), plotted versus the scale length
$\log(h)$ of the detected `disks'. The distribution of values is large: it
goes from small and bright `disks' ($\mu_{0}^{c} < 19$, $h < 0.63$ kpc), many
of them hosted by luminous E galaxies, to less luminous and great disks,
preferably found in S0 and low luminosity ellipticals. The disk parameters of
normal spirals (from \opencite{dejong}) are also included in the diagram for
comparison.

It seems that the whole distribution follows approximately the
lines of constant disk luminosities (dashed lines).
The spiral disks of de Jong's sample are brighter than the early-type disks
of our sample and extend towards fainter $\mu_{0}^{c}$ and longer scale
length $h$. The distribution of $\mu_{0}^{c}$ does not peak around
the Freeman value (solid line)
(\opencite{freeman}, \opencite{kruit}).

In Fig. 4 we plotted the effective radius $\log(R_{e})$ of the bulge component
versus the scale length $\log(h)$. It is apparent there is a correlation
between these parameters, but with two separated trends, one for the
bulge dominated systems (upper part of the diagram), and another for the disk
dominated galaxies (bottom part of the diagram).

It seems also that galaxies with very small bulges and disks ($R_{e} \sim h <
0.6$ kpc) as well as systems with large $R_{e}$ and $h$ ($>2$kpc) do not exist.
We note that the lack of objects with large size cannot be a bias, since our
sample is complete in luminosity.

\section{Conclusions}
By applying a full 2D fit of the light distribution of a volume-limited
sample of early-type galaxies we obtained the following results:

\begin{enumerate}

\item The luminosity distribution of bright E galaxies is well 
represented by the superposition of an inner exponential component and 
a large outer bulge.
This inner structure is characterized by an hight central surface 
brightness and a small scale length.
It is likely that the detected component is the core of the galaxy.
It seems important, however, that this component follows the same photometric 
properties of normal disks, although shifted in zero point.

\item The distribution of $\mu_{0}^{c}$ for all detected `disks' is 
a function of the scale length $h$. The slope of the relation 
is that of constant disk luminosities. The central surface brightness
$\mu_{0}$ spans a wide range of values, in contrast with the 
claimed constant value. 

\item The effective radii of the bulges and the scale lengths of 
the `disks' are correlated: this suggests a possible coupling during
the galaxy formation between the two components, and that the angular momentum
should be an important parameter for the Hubble sequence.

\item The effective radius $R_{e}$ correlates with $\mu_{e}$ and with the 
exponent $n$ of the generalized de Vaucouleurs law. 
The existence of two families of spheroids is 
confirmed as well as the increasing of $n$ for large
luminous bulges. This suggests that elliptical galaxies are not homologous 
systems.

\end{enumerate}

\end{article}
\end{document}